\documentclass[prl,onecolumn,secnumarabic,superscriptaddress,showpacs,
showkeys,amssymb,amsmath,nobibnotes,aps]{revtex4}

\usepackage{graphics}              
\usepackage{latexsym}
\usepackage{bm}                    
\usepackage{amsthm,amsfonts,url}   
\usepackage{verbatim}              

\expandafter\ifx\csname package@font\endcsname\relax\else
 \expandafter\expandafter
 \expandafter\usepackage
 \expandafter\expandafter
 \expandafter{\csname package@font\endcsname}
\fi

\def\blfootnote{\xdef\@thefnmark{}\footnotetext}

\begin{document}

\title{On the logical structure of Bell theorems}


\author{Anne Broadbent}
\email{broadbea@iro.umontreal.ca}
\affiliation{DIRO, Universit\'e de Montr\'eal, C.\,P.~6128,
Succursale Centre-Ville, Montr\'eal, Qu\'ebec, Canada H3C 3J7}

\author{Hilary A. Carteret}
\email{hcartere@qis.ucalgary.ca}
\affiliation{IQIS, University of Calgary, 2500 University Drive NW,
Calgary, Alberta, Canada T2N 1N4}

\author{Andr\'e Allan M\'ethot}
\email{methotan@iro.umontreal.ca}
\affiliation{DIRO, Universit\'e de Montr\'eal,
C.\,P.~6128, Succursale Centre-Ville,
Montr\'eal, Qu\'ebec, Canada H3C 3J7}

\author{Jonathan Walgate}
\email{jwalgate@qis.ucalgary.ca}
\affiliation{IQIS,
University of Calgary,
2500 University Drive NW,
Calgary, Alberta, Canada T2N 1N4}

\date{July 23, 2006}

\begin{abstract}
Bell theorems show how to experimentally falsify local realism. Conclusive falsification is highly desirable as it would provide support for the most profoundly counterintuitive feature of quantum theory---nonlocality. Despite the preponderance of evidence for quantum mechanics, practical limits on detector efficiency and the difficulty of coordinating space-like separated measurements have provided loopholes for a classical worldview; these loopholes have never been simultaneously closed. A number of new experiments have recently been proposed to close both loopholes at once. We show some of these novel designs fail in the most basic way, by not ruling out local hidden variable models, and we provide an explicit classical model to demonstrate this. They share a common flaw, which reveals a basic misunderstanding of how nonlocality proofs work. Given the time and resources now being devoted to such experiments, theoretical clarity is essential. Our explanation is presented in terms of simple logic and should serve to correct misconceptions and avoid future mistakes. We also show a nonlocality proof involving four participants which has interesting theoretical properties.
\end{abstract}

\pacs{03.67.-a, 03.67.Mn}
\maketitle


\section{Introduction}

Some specific predictions of quantum mechanics are inconsistent with
local realism~\cite{bell}. Either these predictions are false or
else our world is not locally realistic. These predictions can be
tested, as quantum mechanics is a physical theory; however they are
hard to verify indisputably. Independent measurements must be made
upon systems that share an entangled quantum state yet which cannot
possibly be dynamically linked, a state of affairs only achieved by
space-like separation. Furthermore, these measurements must be
sufficiently reliable to prevent any hypothetical local conspiracy
from exploiting errors to create an illusory quantum effect.
These
two challenges, respectively the `locality
loophole' \cite{lloop1, lloop2} and the `detection loophole'
\cite{dloop1, dloop2}, have yet to be met in a single
experiment. To dispel this classical paranoia once and for all, we
must find quantum predictions so profoundly divergent from local
realism that the simultaneous closure of these loopholes is
feasible.

A new kind of nonlocality proof has emerged in the recent literature and received widespread attention precisely for its apparent ability to close both loopholes simultaneously; examples are the two-photon experiments proposed by Cabello \cite{cab1,cab2} and by Greenberger, Horne and Zeilinger \cite{ghz1,ghz2}. These proofs simplify matters by reducing the number of different photon measurements required to violate classicality, thereby reducing the detector-efficiency threshold of the detection loophole to feasible levels. These proposals are flawed, however, in the sense that they do not rule out the most general type of local theory, exposing an important misconception concerning the structure of nonlocality proofs. The shortcut they take necessarily introduces an additional assumption into the proof procedure, and however plausible this assumption may be it allows local realism to evade contradiction. Though our argument is based on simple reasoning we are not just splitting logical hairs. This flaw allows local models to pass these `nonlocality tests' with flying colours, as we show by explicit construction. Were such experiments performed with perfect detectors, they would still not falsify local realism. One of the two purposes of this paper is to explain clearly what this increasingly common mistake is, and how not to make it. A prior manuscript by two of the authors on the broader subject of
``Entanglement swapping, light cones, and elements of reality'' has
presented some classical explanations of these proposed violations of
local realism \cite{bm06}. The quantum violations of local
realism are one of the theory's strangest and most perplexing
features. If progress is to be made towards understanding our
fundamentally nonclassical world, it is of paramount importance we
understand precisely the experimental evidence in favour of
nonlocality. This is especially important in light of the considerable resources now being devoted to realizing loophole-free experiments. There is much value therefore in a detailed examination of
the structure of nonlocality proofs, and in exposing a tempting
shortcut as a logical dead end.

The second purpose of this paper is to draw attention to an
as-yet-unnoticed feature of Cabello's proposed experiment. Although
his two-player proof is flawed, we show it can be converted into a
valid four-player pseudo-telepathy game that comprises four nested
three-party Mermin-GHZ games, played simultaneously upon the same
quantum state. One might imagine that a state exhibiting four nested
Mermin-GHZ type correlations would need each party to control more
than one qubit.  We show how to do this with each party holding just
one qubit.

The paper is divided into three parts. In Section~II we set the
scene, introducing the key elements of nonlocality proofs, and recalling the salient features of Cabello's experimental proposal. Though our analysis is general we will focus upon one example for the sake of clarity, and we choose Cabello's design because it is perhaps the most convincing of its class, and has been clearly presented on a number of occasions in the literature \cite{cab1,cab2,cab3}. This experimental model is set in a framework that has the potential to produce `all-versus-nothing' violations of local realism: a non-maximally entangled four-qubit system. In
Section~III, we study
this purportedly `loophole-free' two-party Bell experiment and provide a classical
model that perfectly reproduces all of the observed correlations, showing these proofs are not
valid. We find the flaw lies in an unwarranted assumption about the
nature of `local elements of reality'. Interestingly, although such
assumptions are intuitively reasonable, they are fatal to
nonlocality proofs. We provide a detailed explanation of why this is
so. Finally, in Section~IV we show how a four-party analogue of
the flawed proposal yields a valid pseudo-telepathy game and discuss
its novel properties.

\section{Four Qubit Nonlocality} \label{sec2}

`Bell theorems without inequalities', also called `nonlocality
without inequalities experiments', are promising candidates for a loophole-free local realism falsification \cite{heywood,ghz0,mermin,hardy}. They identify sets of measurements
that, when performed independently upon an entangled system, produce a
set of possible outcomes that is qualitatively different from any set of
possible outcomes from any locally realistic model of the experiment. The discrepancies between the quantum and classical predictions come
in two classes: (1) any locally realistic model will unavoidably
produce outcomes that are never produced by the quantum system, and
(2) no locally realistic model can produce all the outcomes that are
produced by the quantum system~\cite{methot}. Were such an
experiment performed many times with perfect apparatus, the list of
recorded outcomes would quickly convince us whether our experiment
was behaving in a locally realistic fashion or not. Standard
Bell-inequality experiments, in contrast, have no sharp distinction
between the sets of outcomes; rather, it is
the {\emph{frequency}} of certain outcomes that is inexplicable by local
hidden variables.

The first class of
Bell theorems without inequalities, sometimes called
`all-versus-nothing' nonlocality proofs, are known to be
equivalent to pseudo-telepathy games~\cite{methot}. These provide an even
stronger refutation of the local realistic viewpoint than the
second class~\cite{methot,bmt05}. Pseudo-telepathy games can be won all
the time by players who share an entangled state, but players using any
classical strategy instead will lose with a non-zero probability~\cite{brass1,brass2}. The novel experiments designed to close both the locality and the detection
loopholes all attempt to take the form of `all-versus-nothing' nonlocality proofs.

Cabello presents two essentially identical nonlocality proofs in
a four-qubit setting \cite{cab1,cab2}. It is well known that
entangled four-qubit systems can provide violations of local
realism, and this system is no exception. However,
these four qubits are encoded onto a two-photon system. Here we
concisely recall the ingredients.

We consider a four-qubit state prepared upon two photons entangled in
both their polarization $(H,V)$ and their path $(u,d)$ degrees of freedom:
\begin{equation}
|\psi\rangle=
\frac{1}{2}(|Hu\rangle_{A}|Hu\rangle_{B}+|Hd\rangle_{A}|Hd\rangle_{B}
+|Vu\rangle_{A}|Vu\rangle_{B}-|Vd\rangle_{A}|Vd\rangle_{B}).\nonumber
\end{equation}
Rewriting this explicitly as a four-qubit state, we have:
\begin{equation}\label{eq:psistate}
|\psi\rangle=
 \frac{1}{2}(
 |0\rangle_{1}|0\rangle_{2}|0\rangle_{3}|0\rangle_{4}
 +|0\rangle_{1}|1\rangle_{2}|0\rangle_{3}|1\rangle_{4}+
 |1\rangle_{1}|0\rangle_{2}|1\rangle_{3}|0\rangle_{4}
 -|1\rangle_{1}|1\rangle_{2}|1\rangle_{3}|1\rangle_{4}).
\end{equation}
Qubits $1$ and $2$ correspond to the polarization and path of
Alice's photon respectively, and likewise for qubits $3$ and $4$ for Bob.
Now consider the following three measurements $X_j, Y_j$ and $Z_j$,
performed individually on qubits $j$ ($j=1\ldots4$):
\begin{eqnarray}
X_{j} &= | 0 \rangle_{j} \langle 1 |
         + | 1 \rangle_{j} \langle 0 | \nonumber \\
Y_{j} &= i (| 1 \rangle_{j} \langle 0 |
         - | 0 \rangle_{j} \langle 1 |) \label{measure}\\
Z_{j} &= | 0 \rangle_{j} \langle 0 |
         - | 1 \rangle_{j} \langle 1|\,. \nonumber
\end{eqnarray}
Each of these measurements has two possible outcomes which we label
$+1$ and $-1$. Let the outcome of measurement $X_{j}$ be written
$x_{j} \in \{ +1,-1 \}$, and similarly for $Y_j$ and $Z_j$. Quantum
mechanics tells us that when appropriate measurements are made on
state  $|\psi \rangle$, the following fourteen equalities will
always be found to hold:
\begin{align}
z_{1}&= \phantom{-}z_{3}, \label{first} \\
 z_{2} &=\phantom{-}  z_{4}, \\
x_{1}  &=\phantom{-}x_{3}z_{4}, \label{1}\\
 x_{2} &=\phantom{-}z_{3}x_{4}, \\
x_{1}z_{2}&=\phantom{-} x_{3}, \label{14eqns:5} \\
z_{1}x_{2}&= \phantom{-} x_{4}, \\
y_{1} &=-y_{3}z_{4}, \label{2}\\
y_{2} &=-z_{3}y_{4}, \\
y_{1}z_{2}&= -y_{3}, \label{14eqns:9} \\
z_{1}y_{2}&=  -y_{4}, \\
x_{1}x_{2}&=\phantom{-}y_{3}y_{4}, \label{3} \\
x_{1}y_{2}&=\phantom{-}y_{3}x_{4}, \\
y_{1}x_{2}&=\phantom{-}x_{3}y_{4} \quad \text{and} \label{4} \\
y_{1}y_{2}&=\phantom{-}x_{3}x_{4}. \label{last}
\end{align}

There is no way to allot the values $-1$ and $+1$ to the twelve
outcomes $\{ x_{j}, y_{j}, z_{j} \}$  and satisfy all these
equations simultaneously. A subset of just four equations, for
instance \eqref{1},\eqref{2},\eqref{3} and \eqref{4} already leads
to a contradiction. Therefore any physical theory that demands these
values be preassigned before the measurement choices $\{ X_{j},
Y_{j}, Z_{j} \}$ are made is not consistent with quantum mechanics.

This inconsistency can indeed be exploited to obtain an all-versus-nothing
nonlocality proof. We must be careful, however, that the
measurements $\{ X_{j}, Y_{j}, Z_{j} \}$ are performed in such a way
that local realism {\emph{requires}} the values $\{ x_{j}, y_{j}, z_{j} \}$
be preassigned. This is easy to guarantee if the four qubits are
space-like separated, but a complication arises when the four qubit state
$|\psi\rangle$  is instantiated upon Cabello's two-photon system.
Qubits $1$ and $2$, the polarization and the path of Alice's photon,
clearly cannot be measured at space-like separation. The same
clearly applies to Bob's photon, so rather than making four
independent qubit measurements chosen from three alternatives, we are
really making two independent photon measurements chosen from nine
alternatives:
\begin{equation}
\{
(X_{1},X_{2}); (X_{1},Y_{2}); (X_{1},Z_{2}); (Y_{1},X_{2}); (Y_{1},Y_{2});
(Y_{1},Z_{2}); (Z_{1},X_{2}); (Z_{1},Y_{2}); (Z_{1},Z_{2}) \}.
\end{equation}
Cabello permits Alice and Bob to refrain from measuring one of their
qubits, which leads to $9+6=15$ possible local measurements, but
this complication does not affect the analysis. These two
measurements each have four possible outcomes:
$\{(-1,-1);(-1,+1);(+1,-1);(+1,+1)\}$ There is no \emph{logical}
reason to assume that just because $x_{1}=1$ when $(X_{1},X_{2})$ is
measured, $x_{1}$ would have equalled~$1$ if we had measured
$(X_{1},Y_{2})$. Perhaps the different apparatus required to measure
different path observables affects the photon's observed
polarization? If we want to rule out this possibility, we must
design our experiment very carefully. Quantum mechanics may tell us
these measurements are independent, but nothing prevents
local hidden variables from disobeying this rule!


\section{The Logic of Nonlocality Proofs} \label{sec3}

The `nonlocality proof' of Section~II works as follows.
(Cabello's two papers
provide two different descriptions of essentially the same proof; for
ease of reference we discuss only that formulated in \cite{cab1},
but our objection and counterexample apply equally to the
equivalent formulation in \cite{cab2}.)
Alice randomly chooses to perform one of the following two measurements:

\begin{quote}
1a. $X_{1}$ and $X_{2}$? \\
2a. $Y_{1}$ and $X_{2}$?
\end{quote}
Bob meanwhile randomly performs one of the following four measurements:

\begin{quote}
1b. $X_{3}$ and $Y_{4}$? \\
2b. $X_{3}$ and $Z_{4}$? \\
3b. $Y_{3}$ and $Y_{4}$? \\
4b. $Y_{3}$ and $Z_{4}$?
\end{quote}
The only relevant equations are \eqref{1},~\eqref{2},~\eqref{3}
and~\eqref{4}, as none of the other equations are ever tested by this experiment. Quantum mechanics predicts that these equations will always be satisfied. For this to be a valid nonlocality proof, there must be no way for a local hidden variable model to achieve the same thing. Yet the following classical model not only does exactly that, but also manages to perfectly mimic the quantum measurement statistics without even needing to hide the classical variables!

Let $\lambda_{1}$, $\lambda_{2}$ and $\mu$ be three independent random bits taking the values $+1$ or $-1$ with equal probability. These will be the local hidden variables of our classical model. Instead of two entangled photons, Alice and Bob share a two-part system each part of which carries a copy of $\lambda_{1}$, $\lambda_{2}$; Bob also has a copy of $\mu$.

Alice's part of the system behaves as follows---regardless of whether she performs measurement 1a or 2a, it will simply output ``$\lambda_{1}$ and $\lambda_{2}$'':
\begin{quote}
1a. $\rightarrow \ \ \lambda_{1}$ and $\lambda_{2}$. \\
2a. $\rightarrow \ \ \lambda_{1}$ and $\lambda_{2}$.
\end{quote}
Bob's system produces the following measurement outcomes:

\begin{quote}
1b. $\rightarrow \ \ \mu$ and $\mu \lambda_{1} \lambda_{2}$. \\
2b. $\rightarrow \ \ \mu$ and $\mu \lambda_{1}$. \\
3b. $\rightarrow \ \ \mu$ and $\mu \lambda_{1} \lambda_{2}$. \\
4b. $\rightarrow \ \ \mu$ and $- \mu \lambda_{1}$.
\end{quote}

It is easy to see that in perfect agreement with quantum mechanics, the result of each individual `qubit' measurement is completely random, yet the global correlations of equations \eqref{1},~\eqref{2},~\eqref{3} and~\eqref{4} always hold. This local model is, in the context of this experiment, utterly indistinguishable from quantum mechanics itself, and this needs just two shared random bits and one private random bit to achieve. Since the experiment admits such a simple locally realistic explanation, it cannot have falsified local realism!

It is worth noting that while the above local model mimics quantum mechanics, this is not required if we simply want to pass the `nonlocality test' set by the four equations \eqref{1},~\eqref{2},~\eqref{3} and~\eqref{4}. In this case an even simpler solution presents itself, whereby Alice and Bob needn't bother measuring their system at all, but simply respond in the following mechanical fashion:

\begin{tabular}{l l}
Alice: &always say ``+1 and +1'' \\
Bob: &if asked questions 1b, 2b or 3b, say ``+1 and +1'' \\
\quad &if asked question 4b, say ``+1 and $-1$''.
\end{tabular}
\vspace{12pt} \\
These
three lines suffice to prove that the experiment is not a nonlocality proof. For what would an
experiment prove, when we know that the correlations it generated
could be matched by two classical automata running
this communication-free, postage-stamp sized protocol?

\bigskip

It is argued in~\cite{cab1,cab2}, that strategies such as
this are forbidden in the new type of nonlocality proof. In the
proposed model, Bob must always give the same answer to the question
``What is $z_{4}$'', regardless of the context in which that
question is asked: ``\emph{Since $z_{4}$ represents a local element
of reality}, Bob's answer to $Z_{4}$ must be independent on whether
$Z_{4}$ is asked together with $Y_{3}$ or $X_{3}$'' (emphasis
added). This is exactly the misconception at the heart of this and
other recent
proposals for `improved' nonlocality proofs. We must not
make any assumptions about what constitutes a local element of reality!
Any alleged proof that spends any time whatsoever establishing `what the
local elements of reality must be' is likely to be wrong,
or, at the very least, not as general as it should be.

Nonlocality proofs share a simple logical structure: they are proofs
by contradiction. Two assumptions are made---the assumption of
\emph{locality} and the assumption of \emph{realism}. A valid
argument leads from these premises to a conclusion concerning the
possible outcomes of measurements upon causally unconnected systems.
It is then shown that this conclusion is false if the predictions of
quantum mechanics for certain space-like separated entangled states
are true. When these predictions are experimentally verified, the
conclusion is experimentally refuted, and therefore at least one of
our two premises must have been false.

The new model for nonlocality proofs has a different, two-step
structure, which Cabello erroneously attributes to Bell \cite{cab3},
and which is mirrored in the recent proposals by Greenberger, Horne
and Zeilinger \cite{ghz1,ghz2}. In the first step, some predictions
of quantum mechanics for the behaviour of a specific state
$|\psi\rangle$ under a specific set of possible measurements $\{
X_{j}, Y_{j}, Z_{j} \}$ are assumed to be true. To be specific,
in the proposed model, it is concluded that pairs of measurements upon
different qubits encoded on the same photon are outcome
independent; the outcome of measurement $A$ on qubit $1$ is shown to
be independent of the choice of measurement on qubit $2,$ and vice
versa. A valid argument leads from this premise to a preliminary
conclusion concerning the nature of viable local hidden-variable
(LHV) models.
In the second step, locality, realism, \emph{and the conclusion of
the first step} are assumed, and a deduction is made concerning the
possible outcomes of measurements upon causally unconnected systems.
It is then shown that this conclusion is false if some other
predictions of quantum mechanics for the state $|\psi\rangle$ are
true (to be specific, equations \eqref{1},~\eqref{2},~\eqref{3}
and~\eqref{4}).

The problem with this two-stage approach should be apparent. When we
conduct
the experiment presented in Section~II using two photons,
our logical conclusion will be shown to be inconsistent with
observable evidence. We can deduce that at least one of the premises
of our overall argument must have been false.
However, the proposed new type of nonlocality proof has a total of
three premises, not two! In addition to locality and realism, it is
assumed from the outset that \emph{in any LHV model, measurement
outcomes that represent `local elements of reality' (as defined by Einstein, Podolsky and Rosen) must be assigned
definite values}.
The proposed nonlocality proof does not test to see if this
assumption is true for the system and measurements in question. Thus, the ensuing experiment will not rule out
local realism. The third assumption can act as a
`logical shield', protecting locality and realism from
contradiction.
It must also be noted that the term coined by Cabello, `Einstein,
Podolsky, Rosen local elements of reality' or EPRLERs, is
misleading. Einstein, Podolsky and Rosen never put forth a
definition of
a LER but only
offered a criterion to \emph{recognize} one~\cite{epr35};
they explicitly allowed for the possibility of other models.

There is nothing
logically invalid about using three assumptions, instead of just
the two. We certainly don't reject the third premise because we're
forbidden from making spurious and unsupported assumptions about the
properties of reality. After all, the assumptions of locality and
realism are (surprisingly!) poorly physically motivated, whereas
Cabello's additional assumption is experimentally verifiable. At the
end of the day, we can make any assumptions we want, but the conclusion we will end up drawing is that `one of our assumptions
must be wrong'. If we want to rule out local realism, we'd better
not have any additional assumptions in the way that can act as
sacrificial pawns. If we have assumed some quantum predictions \emph{without testing them},
logic dictates that these predictions might be wrong, however
reasonable they seem. In this case, the application of logic may
appear physically counterintuitive: an implicit assumption that quantum
mechanics describes what is really physically happening
leads to a proof with a classical solution! Nevertheless the logic
is indisputable: the classical model is extremely simple and perfectly reproduces the supposedly nonlocal quantum correlations; an
experiment with a classical explanation cannot prove
nonclassicality. This highlights the value of proper logical analysis. The existence of an additional necessary assumption can be used as a test for the possibility of a local hidden variable solution, saving one the effort of exhaustively construction new local models of every specific case.

It is important to be very clear about our reasons for rejecting the
additional assumption, so let us reiterate one last time. It is fatal to include an additional assumption in nonlocality proofs, even if that assumption is known to be \emph{true} for quantum mechanics, because doing so can open the door to LHV models for which that additional assumption is \emph{false}. Cabello's errant
assumption is surely true,
as it is a mathematical property of quantum mechanics.
Nevertheless,
when the validity of quantum mechanics itself is at issue,
it is a mistake to \emph{assume} it and not
\emph{test} it, as must be clear from the simple
counterexample presented above---quantum assumptions have led to a classical solution.

It is of course a
physically observable fact that the measurement $X_{1}$ produces outcomes that are
independent of the measurement performed on qubit~$2$, and it is
quite possible to show that quantum mechanically it must be so. Why can we not perform this experiment separately, prove the errant assumption to our satisfaction, and then combine the results? Because doing so would introduce loopholes that dwarf those we are attempting to close. How could we claim to have closed the locality loophole, for instance, unless we performed the two experiments themselves at spacelike-separated locations? More damningly, how could we rule out a local physics that was sensitive to the differences between our experimental setups? Nonlocality proofs aim not just to meekly persuade us of the violation of local realism, but to logically compel us to accept it. This is their great strength. Confronted with the profoundly counterintuitive phenomenon of quantum nonlocality, it is vital to establish its reality as firmly as we can. Bell's profound result was that we could falsify the entirety of local realism with one tantalisingly realizable experimental setup. But only if we do that experiment right!


How do we do things right? We must get rid of
the additional assumption. We can redesign our experiment such that
in parallel with everything else, it actually tests whether
\emph{all} the predicted behaviours of the quantum state
$|\psi\rangle$ under measurements $\{ X_{j}, Y_{j}, Z_{j} \}$ are
observed, both equations (\ref{1}),(\ref{2}),(\ref{3}),(\ref{4})
and the independence of separate qubit measurements. This revision
guarantees the only assumptions that might be false are locality and
realism. Testing additional predictions means we will have to ask
Alice and Bob to perform some additional measurements. It is exactly these
measurements that Cabello adds to his original experiment in order to create
a valid nonlocality proof in his recent response to criticism
\cite{cab3}.
(Of course the validity of the extended experiment was never
questioned, and does not imply the validity of the original smaller
experiment, just as an attempt to test the Clauser, Horne, Shimony and Holt inequality that performed measurements in only one of the two nonorthogonal bases would not violate local realism. Half a valid proof is no proof at all.)
However, the original
proposal explicitly avoided testing these additional predictions in
order to reduce the supposed maximum classical success rate
to~$\frac{3}{4}$ and thereby ease the burden placed on the photon
detectors. As we have shown, this was unsuccessful. The valid
extended experiment works because it tests all fourteen equations \eqref{first} to~\eqref{last}. A local hidden variable model can reproduce these correlations with probability at most~$\frac{13}{14}$, significantly worse than other two-party proposals~\cite{aravind02},
and thus is not ``stronger'' or ``loophole-free'' in any meaningful
sense.

\bigskip

There is a different way to make the original experiment valid. We can
abandon the two-photon instantiation of $|\psi\rangle$ and consider
four space-like separated qubits. The resulting experiment does not yield
a better experimental proposal than the Mermin-GHZ
pseudotelepathy game~\cite{ghz0,mermin} if we are concerned with closing the
detection loophole or with minimizing the number of participants,
but nevertheless has a number of interesting properties that we
discuss in the next section.


\section{The Four-Party Pseudo-Telepathy Game} \label{sec4}

As mentioned in Section~II, if the variables of
equations~(\ref{first})--(\ref{last}) are to be taken as
independent, then it is impossible to assign predetermined values to
the variables in order to satisfy all these equations. An easy way
to guarantee independence is to have all the measurements on the
different qubits performed in space-like separated regions: in the
language of pseudo-telepathy, give each qubit to different
participants. Therefore, if we see the nonlocality argument based on
equations~(\ref{first})--(\ref{last}) as a four party experimental
setup, we do have a valid refutation of local realism.

It is interesting to mention that the subset of four equations,
Equations~(\ref{1}), (\ref{2}), (\ref{3}) and~(\ref{4}), bears a
striking resemblance to Mermin-GHZ's equations in the three-party
setting~\cite{ghz0,mermin,brass1}, where the second player is not
`involved'. His output basically decides which Mermin-GHZ type game
the other three participants are playing (see equations
(\ref{1a})--(\ref{4b})). Furthermore, many different other
combinations of equations will give the same scenario, with
different participants deciding which game the others are playing.
Since the Mermin-GHZ correlations are enough to refute the local
realistic viewpoint, one might wonder why we should bother with the
fourth player, the third measurement setting and the added number of
different measurement setups of our scheme. From the perspective of
closing the detection loophole, as far as we know, there is no
justifiable reason. However, we would like to point out that this
new pseudo-telepathy game does not require a GHZ state, but the
$\vert \psi \rangle$ state. Therefore, under certain experimental
situations, those where a GHZ state is difficult to create or
manipulate, our scheme could be advantageous over the traditional
Mermin-GHZ scheme.

We can see our game as four embedded three-player Mermin-GHZ games,
each one having a different participant that is not involved.
Picture the scenario as following: a referee designates one of the
four players, say Bob, as a co-referee. Bob now has to choose
whether Alice, Charlie or Didier will play the first Mermin-GHZ
game:
\begin{align}
x_{1} &=\phantom{-}x_{3}z_{4}, \label{1a}\\
y_{1} &=-y_{3}z_{4} \label{2a},\\
x_{1} &=\phantom{-}y_{3}y_{4} \quad \text{and} \label{3a} \\
y_{1} &=\phantom{-}x_{3}y_{4}, \label{4a}
\end{align}
\noindent or whether they will play the second Mermin-GHZ game:
\begin{align}
x_{1} &=\phantom{-}x_{3}z_{4}, \label{1b}\\
y_{1} &=-y_{3}z_{4}, \label{2b}\\
x_{1} &=-y_{3}y_{4}\quad \text{and} \label{3b} \\
y_{1} &=-x_{3}y_{4}. \label{4b}
\end{align}
\noindent It must be noted that the players do not need to know who
was chosen as co-referee. Even the co-referee himself can be left in
the dark concerning his role. They are simply asked questions
which they must answer; the division lays in the head of the
referee. It is also interesting to note that a similar game can be played
with the players sharing a four-party GHZ state $(\vert 0000 \rangle
+ \vert 1111 \rangle )/\sqrt{2}$. However, it is surprising that
this GHZ state is \emph{not necessary}. Furthermore, one might be
tempted to think that four three-party GHZ states are necessary, a
total of twelve qubits, even if the game is not really four times
the Mermin-GHZ game, while we can play this game with a low-dimensional
state. It would also seem natural that
Mermin-GHZ correlations could only arise with GHZ states. Our scheme
demonstrates that it is not the case and that the $\vert \psi
\rangle$ state is sufficient.
This is rather surprising, because this state is not equivalent
to the four-qubit GHZ state $(|0000\rangle + |1111\rangle)/\sqrt{2}$
under local unitary operations.
It is interesting to point out that our game is reminiscent of the
game presented in~\cite{av}. However, the game of~\cite{av} is
presented as a game where players collude in order to form a
standard Mermin-GHZ game.


\section{Conclusion}

Nonlocality is one of the most mysterious aspects of quantum
mechanics and it is difficult to grasp for us living in an
apparently classical world. Nonlocality proofs give us very useful
tools to not only demonstrate nonlocality in theory, but also to
implement convincing experiments in the laboratory. We have shown
that a conceptual error in the design of nonlocality proofs can be
fatal to the ultimate goal of such a proof, which is to demonstrate
that our world is not locally realistic.

More precisely, we have elucidated why the description of a good
nonlocality proof can (and should) be given \emph{without any
discussion of quantum mechanics or the nature of local elements of
reality}. It is only the actual experimental setup, or the quantum
winning strategy that needs to invoke quantum mechanics. We have
shown that doing otherwise can fatally compromise the conclusions
that can be drawn from nonlocality proofs. Because the
two-participant nonlocality proofs of Cabello \cite{cab1,cab2} need
to invoke properties of quantum mechanics as assumptions, we conclude that these proposals do not achieve
their purported goal of ruling out locally realistic descriptions of
our world,
in spite of the fact that they do rule out some subclass of LHV
models.

We also have shown that it is possible to modify Cabello's argument
to make it a valid nonlocality proof by using a different method to
that in~\cite{cab3}: by space-like separating the
assumed-to-be-independent measurements, and thus enforcing the
necessary measurement independence. This new construction leads to a
pseudo-telepathy game with a most surprising feature. The proposed
game shows that multiple Mermin-GHZ type correlations can be
obtained from states with parties that control only low-dimensional
non-maximally entangled systems.

\bigskip

\bigskip

We would like to thank Ad\'an Cabello,
Shan Lu,
Barry Sanders, Valerio Scarani and Rob Spekkens for stimulating
discussions. A.\,B. is supported by a scholarship from Canada's
Natural Sciences and Engineering Research Council (NSERC), H.\,C. is
supported by iCORE and J.\,W. is supported by the Alberta Ingenuity
Fund (AIF) and the Pacific Institute for the Mathematical Sciences
(PIMS).  A.\,A.\,M. acknowledges support and hospitality from
Calgary's Institute for Quantum Information Science.

\end{document}